\documentstyle[aps,preprint]{revtex} % REVTeX

\begin{document}

\title { Temperature crossovers in  cuprates}
\author{Andrey V Chubukov\ddag\footnote[1]{Also at:
P L Kapitza Institute for Physical Problems, Moscow, Russia},
David Pines\S\ and Branko P Stojkovi\'c\S\ }
\address{\ddag\ Department of Physics, University of Wisconsin,
Madison, WI 53706}
\address{\S\ Department of Physics, University of Illinios,
Urbana, IL 61801}
\date{\today}

\maketitle % comment this line when using J. Phys. macros

\begin{abstract}
We  study the temperature crossovers seen in the magnetic
and transport properties of cuprates using a
nearly antiferromagnetic Fermi liquid model (NAFLM).
We distinguish between underdoped and overdoped systems on the basis of
their low frequency magnetic behavior and so classify the
optimally doped cuprates as a special case of the underdoped cuprates.
For the
overdoped cuprates, we find, in agreement with earlier work, mean-field
$z=2$ behavior of the magnetic variables associated with the fact that
the damping rate of their spin fluctuations is essentially independent
of temperature, while the resistivity exhibits a crossover from Fermi
liquid behavior at low temperature to linear-in-T above a certain
temperature $T_0$.
We demonstrate that above $T_0$ the proximity of the quasiparticle
Fermi surface to the magnetic Brillouin zone boundary brings about the
measured linear in $T$ resistivity. For the underdoped cuprates we
argue that the sequence of crossovers identified by Barzykin and
Pines in the low frequency magnetic behavior (from mean field $z=2$ at
high temperatures, $T>T_{cr}$, to non-universal $z=1$ scaling
behavior at intermediate temperatures, $T_*<T<T_{cr}$, to pseudogap
behavior below $T_*$) reflects the development in the electronic
structure of  a precursor to a spin-density-wave state. This
development begins at $T_{cr}$
%, identified by Barzykin and Pines
%  (1995) as the temperature at which $\xi\sim2a$,
with a thermal evolution of the quasiparticle spectral weight
which brings about temperature dependent spin-damping and ends at $T_*$
where the Fermi surface has lost pieces near corners of the magnetic
Brillouin zone. For $T_*<T<T_{cr}$ the resistivity is
linear in $T$ because this change in spectral weight does not affect the
resistivity significantly; below $T_*$ vertex corrections act to 
bring about the
measured downturn in $(\rho(T)-\rho(0))/T$ and approximately
quadratic in $T$ resistivity for $T\ll T_*$.
\end{abstract}


\section{Introduction}
\label{sec:intro}

Over the past few years, it has become increasingly clear to the
``high-$T_c$''
community that the mechanism of superconductivity in cuprates is
directly
related to their unusual properties in the normal state, particularly in
the underdoped regime.
The $^{63}$Cu spin-lattice relaxation rate and spin-echo decay rate
(Slichter 1994, Barzykin and Pines 1995), uniform
susceptibility (Johnson 1989), in-plane and c-axis resistivity,
$\rho_{xx}$ and
$\rho_c$ (Ong 1990, Iye 1992),
all demonstrate temperature dependences which in a wide
temperature
range are different from the predictions of Landau Fermi liquid theory.
The remarkable sequence of crossovers (from non-universal
mean field behavior with a dynamical exponent $z=2$
 to $z=1$ pseudoscaling behavior to pseudogap
behavior) seen in magnetic experiments in the normal state of the
optimally doped and underdoped cuprates are shown in Figure
\ref{fig:6}.
Characterizing and explaining this behavior, which possesses
counterparts in angle-resolved photoemission experiments
(Shen et al 1996, LaRosa et al 1996, Ding et al 1996),
transport measurements (Hwang et al 1994),
and optical experiments (Puchkov 1996) on the underdoped cuprates,
is perhaps the major challenge presently facing the high-T$_c$
community.

The effectively non-Landau-liquid behavior at intermediate energy scales
has
stimulated intensive  discussions on a possible violation of
Fermi-liquid
theory in cuprates
(Anderson 1994).  There is currently no consensus on whether
Fermi-liquid behavior is actually broken at $T=0$.
Some researchers believe that the ground
state of underdoped and optimally doped cuprates is not a Fermi liquid:
in
particular, a non-Fermi liquid ground state is a point of departure for
gauge theories based on spin-charge separation (see, e.g., Lee
and Nagaosa (1992),
Altshuler, Ioffe and Millis (1995)),
and for theories in which
pairing is due to pair hopping between adjacent layers (Chakravarty and
Anderson 1994). Another
conjecture is that there exists an isolated  zero-temperature
quantum
critical point at a doping concentration near optimal doping (Varma
1994). On both sides of
this point, it is proposed that
the system behaves as a Fermi-liquid  at the lowest $T$, but
possesses a non-Fermi-liquid quantum-critical behavior above the
crossover temperature, which near optimal doping is assumed
to be smaller than $T_c$.

The authors of the present paper have argued  in favor of a third
possibility, a
nearly antiferromagnetic Fermi-liquid model (NAFLM) for cuprates (for a
review see, e.g, Pines (1995)).
In this approach there is no spin-charge separation. One assumes
that Fermi-liquid behavior is not destroyed by fluctuations at any
doping
concentration, while the measured anomalous (non-Landau-like)
spin and charge behavior seen in the  normal state properties of the
cuprates arises from a magnetic interaction between
planar quasiparticles which reflects the close approach of even an
optimally-doped system to antiferromagnetism. Initial
support for the NAFLM came from the
NMR experiments which  clearly demonstrate the  difference between Cu
and O spin-lattice relaxation rates and establish the presence of
strong temperature dependent antiferromagnetic correlations
(for a review see Slichter (1994)).
More recently the model has been shown to be also consistent with
neutron scattering data which show that even at optimal doping, the spin
fluctuations are peaked at a wavevector ${\bf Q}$  which is at or near
$(\pi,\pi)$ with a magnitude, $\chi_{\bf Q} \gg \chi_0$, and a half
width, $1/\xi$, which at low temperatures is considerably smaller than
the inverse lattice spacing (Mason et al 1996,
Bourges et al 1996,  Zha, Barzykin and Pines 1996).
Moreover, recent experiments have shown that even at optimal doping
there still exist  propagating spin-waves at  energies
comparable to the exchange integral (Hayden et al 1995).

In this paper, we consider the extent to which
the NAFLM can explain the temperature crossovers
in magnetic and resistivity measurements in
both overdoped  and underdoped cuprates. We will show that the
NAFLM yields a sequence of crossovers and changes in the uniform
susceptibility, NMR relaxation rates and resistivity which are
consistent with experiments. In particular, we will show
how for the underdoped cuprates the Fermi surface
evolution which accompanies the development of a precursor to a
spin-density-wave state gives rise to $z=1$ scaling at intermediate
temepratures, and the pseudogap behavior at the lowest temperatures.

The paper is organized as follows: in the next section we review the
NAFLM description of cuprates and its relation to the underlying
microscopic
models with fermion-fermion interaction.
In Secs.\ \ref{sec:overdoped},
\ref{sec:underdoped} and \ref{sec:optimal}
we discuss the physics
of overdoped, underdoped and optimally doped cuprates, respectively.
Finally, in Sec.\  \ref{sec:conclusion} we summarize our conclusions.

\section{The Nearly Antiferromagnetic Fermi Liquid Model}
\label{sec:naflm}

The canonical NAFLM model
is in some respects a two-fluid model for cuprates: it is equivalent
to assuming
that there independently exist fermionic carriers {\it
and} localized  spins  whose susceptibility is
determined from fits to NMR experiments and is an input parameter in the
theory.
Fermions and spins are coupled by
\begin{equation}
{\cal H}_{int} =
\sum_{{\bf q,k},\alpha,\beta} g_{\bf q}\,
c^{\dagger}_{{\bf k+ q}, \alpha}\,
{\vec \sigma}_{\alpha,\beta}\, c_{{\bf k},\beta} \cdot {\vec S}_{\bf q}
\label{intham}
\end{equation}
where $g_{\bf {q}}$ is the momentum-dependent
coupling constant, and $\sigma_i$ are the Pauli matrices.
The momentum dependence of the coupling constant
is not relevant for our considerations, and
for  simplicity we will neglect it throughout the paper.

To second order in the coupling, the spin-fermion interaction
gives rise to an effective pairing interaction between
planar quasiparticles
\begin{equation}
V_{eff}({\bf q},\omega) = g^2\chi({\bf q},\omega)
\label{eff}
\end{equation}
where $\chi(\bf {q},\omega)$ is the susceptibility of localized spins.
NMR and neutron
scattering experiments clearly indicate that  near optimal doping
the spin
fluctuations are overdamped at low energies with a dynamical structure
factor peaked at a wave vector ${\bf Q}$ which is close to
$(\pi,\pi)$ and symmetry related points.
From general considerations, one can then write near ${\bf q}={\bf Q}$
and $\omega =0$ (Millis, Monien and Pines 1990, Barzykin et al 1993,
Monthoux and Pines, 1994b):
\begin{equation}
\chi({\bf {q}}, \omega) =  \frac{\chi_{\bf Q}}{1 + ({\bf q} - {\bf Q})^2
\xi^2 - i
\omega/\omega_{sf} -
\omega^2\xi^2/c_{sw}^2 }
\label{chi}
\end{equation}
where $\chi_{\bf Q} = \alpha \xi^{2}$, $\xi$ is the correlation length,
$c_{sw}$
is the spin-wave velocity,
 and $\omega_{sf} = c^2_{sw}/2\xi^{2}\gamma$ where
$\gamma$ is a damping rate.
The fits to the NMR data show that the parameters, $\xi$, $\chi_{\bf Q}$
and $\omega_{sf}$
depend on temperature, while the scale factor $\alpha$ and the spin-wave
velocity  $c_{sw}$ are virtually independent of T in the temperature
range of NMR experiments (Barzykin and Pines 1995).

It is also
essential that the fermions are {\it not} assumed to be free particles
at
$g =0$, and their Green's function is well defined only near the Fermi
surface
where $G(k,\omega) = Z_k/(\omega - \epsilon_k)$.  The dispersion,
$\epsilon_k$, is
another input parameter in the theory. It is generally assumed to have
the
same
form as in the tight-binding model with hopping between nearest and
next-nearest
neighbors on a square lattice:
\begin{equation}
\epsilon_k = -2 t(\cos{k_x} + \cos{k_y}) - 4
t^{\prime}\cos{k_x}\cos{k_y}
\label{epsilon}
\end{equation}
This form of $\epsilon_k$ (with $t^{\prime} <0$) is chosen to be
consistent
with the shape of the Fermi surface measured in photoemission
experiments
at
around optimal doping (Si et al 1993).
Explicit  calculations show (Monthoux and Pines 1993)
that at these doping levels,
the interaction with spin fluctuations does not  modify
substantially the shape of the Fermi surface as long as the coupling
constant remains roughly smaller than the quasiparticle damping rate.

In principle, the correlation length in (\ref{chi})
should be large enough compared to the interatomic spacing to
ensure that the susceptibility is peaked at ${\bf Q}$.
In practice (Pines 1995), it turns out
that a  correlation length of the order of a lattice spacing already
yields an
appreciable peak in $\chi (\bf {q},\omega)$. Furthermore,
the analytical expansion in
$({\bf q}-{\bf Q})^2$ and $\omega$  is valid only outside the
fluctuation
region for a magnetic phase transition.  This region is however rather
narrow
and most experiments are performed outside it. A more essential point
is that $\chi({\bf q},\omega)$ in Eq.\ (\ref{chi})
does not satisfy the sum rule for the local susceptibility, as the
$3D$ integral $\int d^2 q d\omega \chi({\bf q}, \omega)$ diverges at
the
upper limit. One way to obtain convergence is to include higher-order
 terms in the expansion over momentum. Another way, which we will adopt
here,  is to
introduce a sharp cutoff, $C$, in the integration over
$c_{sw} (\bf {q}-\bf {Q})$. Due to the thermal dependence of $\xi$ and
$\omega_{sf}$ in
Eq.\ (\ref{chi}), the cutoff parameter $C$ also depends on
temperature. On
general grounds, the larger the correlation length is, the smaller is
the
cutoff
scale. However,
$C$ does not vanish when the correlation length diverges.

Although the phenomenological NAFLM considers localized spins as an
independent
degree of freedom, Eqs.\ (\ref{intham}) and (\ref{eff}) can be derived
from
microscopic considerations departing from, e.g.,
the one-band Hubbard model which contains only fermionic degrees of
freedom (Shraiman and Siggia 1988, Schrieffer, Wen and Zhang 1989,
Chubukov and Frenkel 1992,
Bulut, Scalapino and White 1993, Scalapino 1994,
Kampf 1995, Sachdev, Chubukov and Sokol 1995).
In this approach, spin fluctuations appear as a collective mode of
fermions.
To obtain the coupling between two fermions and one spin fluctuation, as
in Eq.\ (\ref{intham}),
one has to dress the original
four-fermion Hubbard interaction term by summing the
RPA series in the particle-hole channel.
As a result of the summation, the product of the two fermionic Green's
functions is replaced by a spin susceptibility whose poles correspond to
spin fluctuation modes. The susceptibility thus obtained
takes the general form
\begin{equation}
\chi({\bf q},\omega) = \frac{\tilde{\chi}({\bf q},\omega)}{1-g
\tilde{\chi}({\bf q},\omega)}
\label{chiq}
\end{equation}
in which $\tilde{\chi}(\bf {q},\omega)$ is the irreducible particle-hole
susceptibility, and $g$ (the same as in Eq.\ (\ref{intham})) is
a coupling constant for the pair interaction between fermions.
For example, for on-site Hubbard interaction, $g = U$.
The momentum and frequency dependence  of the spin
susceptibility obtained in the Hubbard-based approach is generally
consistent
with the phenomenological predictions except that at  any finite doping
level the peak in the susceptibility is located at an
incommensurate
momentum ${\bf Q}$
determined by the maximum in
$\tilde{\chi}({\bf q},0)$. (see, e.g. Gr\"uner 1994).
It is essential however that for fermionic
dispersion as
in Eq.\ (\ref{epsilon}),
the enhancement of
$\chi(\bf {q},\omega)$ comes solely from near resonance in
the
denominator in Eq.\ (\ref{chiq}), while the irreducible particle-hole
susceptibility
does not contain any information about strong magnetic fluctuations,
and near the peak at ${\bf q=Q}$ has a conventional Fermi-liquid form
\begin{equation}
\tilde{\chi}({\bf Q},\omega) =
 \frac{\alpha \tilde{\xi}^2}{1 - i
\omega/\tilde \Gamma_{\bf Q}}
\label{eq:chi_fl}
\end{equation}
where $\tilde \xi$ is of the order of the lattice spacing, and
${\tilde \Gamma}_{Q}$ is
an energy comparable to a fermionic bandwidth. We also assumed
 that there exist
points on the Fermi surface which can be connected by ${\bf Q}$; otherwise,
$\tilde \Gamma_{\bf Q}$ would be infinite.
 It then follows from Eq.\ (\ref{chiq}) that
\begin{equation}
\omega_{sf} \equiv {c_{sw}^2 \over 2\gamma\xi^2} = {\tilde{\Gamma}_{\bf
{Q}}\tilde{\xi}^2 \over \xi^2}
\label{sf}
\end{equation}
We see therefore that the key feature which makes
the physics in the nearly-antiferromagnetic Fermi
liquid different from that in a
conventional Fermi liquid is the presence of an energy scale
 $\omega_{sf} \propto \xi^{-2}$, which vanishes when magnetic
correlation
length becomes infinite, and is  obviously much smaller than the
fermionic
bandwidth as long as the susceptibility is strongly enhanced near
${\bf Q}$.

There are, however, several subtleties with the
microscopic derivation of Eq.\ (\ref{chi}). First, in a RPA
calculation, the fermions which contribute to the spin susceptibility
are generally assumed to behave as free particles everywhere in momentum
space, i.e., one completely  neglects the incoherent part of the
quasiparticle
Green's function. For the calculations of the imaginary part of
$\chi (\bf {q},\omega)$, this assumption  is justified as the integration
over fermionic momentum  is confined to a region where both fermions
in the bubble have momenta near the Fermi surface. For these fermions, the
incoherent part of the fermionic Green's function is irrelevant.
The special points in momentum
space  for which both ${\bf k}$ and
${\bf k+Q}$ are on the Fermi surface are usually referred to as hot
spots (Hlubina and Rice 1995).
The real part of $\chi$, on the other
hand, comes from an integration over regions in momentum space which
are far from the Fermi surface (see, e.g., Rickayzen 1980).
In these regions, the  incoherent part of
$G(k,\omega)$  cannot be neglected as demonstrated in photoemission
experiments (Wells et al 1995, Campuzano et al 1994).
It is therefore likely  that the actual position of the peak in
$Re \chi (q,\omega)$ is different from the RPA result and
may well depend not only on  doping but also on temperature.

Second, even  though the
computation of $Im \chi (q,\omega)$ involves only coherent parts of the
quasiparticle Green functions,
 the RPA approach
neglects possible strong vertex corrections to the polarization bubble.
Explicit calculations show that these corrections are the strongest for
the hot spots (Chubukov 1995, Altshuler, Ioffe and Millis 1995,
Amin and Stamp 1996, Monthoux 1996). At small coupling they are
obviously small to the extent that
$g/\sqrt{\gamma\omega_{sf}}\leq 1$.
At somewhat larger couplings, $g \geq \sqrt{\gamma\omega_{sf}}$,
the relative vertex corrections which come from the
integration near the
Fermi surface scale as $[g^2 \omega_{sf}
\chi_{Q}/v^2_{F}] \log (C/\omega_{sf})$,
 where $v_{F}$ is the Fermi velocity (we set the interatomic
spacing $a=1$).
 Now, if the damping term is computed self-consistently,
then $\omega_{sf}$ by itself scales as $g^{-2}$, and the relative
vertex correction depends on $g$ only logarithmically.
More explicitly,
we have $g^{eff} = g[1 + 2\beta \log (C/\omega_{sf})]$
where $\beta$ depends
only on the shape of the Fermi surface near the hot spots, and is about
$1/16$ for
the experimentally measured  Fermi surface at optimal doping
(Altshuler, Ioffe and Millis 1995, Chubukov 1995).  Notice that
vertex corrections {\it increase}
the spin-fermion interaction ($g^{eff} > g$) and hence
act in favor of magnetically-induced, Eliashberg-type $d-$wave
superconductivity.
We see that $\beta$ is  small numerically, so that if the logarithm is not
large, the relative vertex correction can be neglected.
  In this case, the RPA analysis is approximately valid, and the imaginary
part of the full susceptibility  comes solely from the
imaginary part of the particle-hole
bubble. This is what we believe happens in
the overdoped regime, which we associate with moderate values of
$g \sim \sqrt{\gamma\omega_{sf}}$  in Eq.\ (\ref{intham}).
 However, we will argue
below that the underdoped cuprates are described by a
spin-fermion model with a somewhat larger ratio
$g/\sqrt{\gamma\omega_{sf}}$, where new physics associated
with Fermi surface evolution appears.
If we formally keep using the self-consistent approach,
we find that at the onset of the Fermi surface evolution,
 the logarithmic term overshadows the smallness of
$\beta$,  and the relative vertex
correction  is not small.
In particular, for $g = g^{(1)}_{cr}$ defined below,
$2\beta \log {C/\omega_{sf}} \approx \pi/4$.
In this situation, the validity of the
self-consistent RPA calculations of the damping term also becomes
questionable.

From the above considerations we see that, although some properties of
the
NAFLM can be derived in Hubbard-based calculations, one has to be
careful about using an RPA formalism in calculating the
dynamical spin susceptibility.
We therefore advocate a semi-phenomenological
approach in which it is assumed that the spin susceptibility has the
form of
Eq.\ (\ref{chi}), with both $\xi$ and $\gamma$ taken from fits to
the experimental data. We emphasize however that from
a physical perspective, only $\xi$ should be considered
as an independent input parameter for
Eq.\ (\ref{intham}). The damping of spin excitations due to
the interaction
with fermions is fully described by a spin-fermion model, and we  will use
the phenomenological form for $\omega_{sf}$  only because we are
currently unable to self-consistently compute
 the spin damping at $g \gg c_{sw} \xi^{-1}$. We will, however, be
able to estimate the value of the spin damping at very large couplings
where the electronic structure develops the precursor of
a spin-density-wave form.

We now consider the extent to which the physical properties of the
normal state can be explained by the NAFLM.

\section{Overdoped Cuprates}
\label{sec:overdoped}

We first discuss the overdoped cuprates. We define overdoped systems as
those whose magnetic behavior is such that in the normal state their
uniform magnetic susceptibility, $\chi_0$, is either independent of $T$
or even increases with decreasing temperature, while the product
$^{63}T_1T$ and the square of the $^{63}$Cu spin-echo decay
rate $(^{63}T_{2G})^2$ are linear in $T$.
Examples of overdamped systems are La$_{2-x}$Sr$_x$CuO$_4$ for $x\geq
0.2$ and Bi 2212 and Tl 2201 for the appropriate oxygen content. The
measured resistivity of overdoped cuprates is linear in $T$ above some
particular temperature $T=T_0$ and has a Fermi liquid form $\rho\sim
T^2$ at $T\ll T_0$.

Our theoretical interpretation of the experiments on overdoped cuprates
is based on the conjecture, first put forward by Barzykin and Pines
(1995), that in these systems
$\gamma$ and $c_{sw}$ are independent of temperature,
while $\xi\leq 2$ for all $T>T_c$. This in turn
implies that for overdoped cuprates
\begin{equation}
\omega_{sf} \xi^2=const.
\end{equation}
Since the correlation length $\xi$ (which, we remind the reader, is an
input parameter for NAFLM) does not exceed a few interatomic spacings,
its temperature dependence is governed by fluctuations at lattice
scales and is likely to be material dependent. NMR experiments
tell us that for all compounds studied
$\xi^{-2}$ scales linearly with $T$, i.e., $\xi^{-2}=A+BT$,
where $A$ and $B$ are (material dependent) constants. More
specifically, fits to the NMR data using Eq.\ (\ref{chi}) and the
standard Shastry-Mila-Rice
Hamiltonian for hyperfine interactions (Shastry 1989,
Mila and Rice 1989), then yield (Thelen and Pines 1994):
\begin{equation}
^{63}T_1T \sim {\omega_{SF} \over \alpha}\sim A+BT,
\label{63Ta}
\end{equation}
and
\begin{equation}
(^{63}T_{2G})^2 \sim \left({1 \over \alpha\xi}\right)^2\sim A+BT
\label{63Tb}
\end{equation}
in agreement with the experimental results.
The form of the susceptibility with $\gamma$ and $c_{sw}$ independent
of temperature
and $\xi^{-2}=A+BT$ is formally equivalent to that found in the $z=2$
quantum-critical regime (Sachdev, Chubukov and Sokol
1995). Clearly then, the $T$ dependence of all observables associated
with the spin susceptibility near $(\pi,\pi)$ should be the same as for
$z=2$ scaling (Millis, Monien and Pines, 1990).
In particular,
 the measured ratio $^{63}T_1 T/^{63}T_{2G}^2 \propto \omega_{sf} \xi^2$
is independent of temperature.

However,
the analogy with $z=2$ scaling is only a formal one. Although
it works for observables, such as $^{63}T_1$ and $T_{2G}$,
 which probe the susceptibility near $(\pi,\pi)$, the
behavior of the uniform susceptibility does not follow the
scaling prediction, which is based on the  assumption that spin
fluctuations near ${\bf q}=0$ and ${\bf q}={\bf Q}$
are related. These are not related for the overdoped cuprates because
their physics is dominated by the short wavelength fluctuations.
Indeed,  above $T_{cr}$, $\chi_0 (T)$ decreases with
 increasing $T$ (NMR studies show that
$T_{cr}$ roughly corresponds to the temperature where $\chi_0 (T)$ is a
maximum). On the other hand,
 $1/N$ quantum-critical calculations yield $\chi_0 (T)$ linearly increasing
with $T$ in the  $z=2$ regime (Sachdev, Chubukov and Sokol 1995)
\footnote{There is some uncertainty in the scaling prediction
for $\chi_0 (T)$
 because $d=2$ is an upper critical dimension for the $T=0$
transition. Ioffe and Millis argued that the linear term in susceptibility
could be of either sign (Ioffe and Millis 1995). The  sigma-model based
calculations yield a positive linear in $T$ term (Sachdev, Chubukov and Sokol
1995).}
As microscopic self-consistent
(Eliashberg) calculations show (Monthoux and Pines 1993, 1994a),
the  decrease of susceptibility with increasing temperature
is a natural consequence of
lifetime effects, which act to reduce $\chi_0(T)$,
 and which play an increasingly important role as the
temperature increases.

The behavior of the resistivity $\rho$
is also nonuniversal, and, in principle,
depends on the details of the electronic band structure. However, as we
now show, in an argument which we shall see applies to the optimally
doped cuprates as well as overdoped
cuprates, $\rho (T)$ is in fact proportional to $T$ in a wide
temperature range
due to the proximity of the
quasiparticle Fermi surface to the magnetic Brillouin
zone boundary. The reason is the following:
according to the Drude formula, which is applicable to the overdoped
cuprates because vertex corrections are small,
 the conductivity is
proportional  to the quasiparticle
relaxation time averaged over the Fermi surface.
The inverse relaxation time - the relaxation rate -
is given by the imaginary part
of the quasiparticle self-energy.  As discussed above, the
overdoped regime is associated with small to moderate values of the coupling
constant. In this situation, second-order perturbation theory is
likely to be valid. To second order in $g$
we find by  performing a summation over frequencies:
\begin{equation}
{1\over \tau_{\bf k}} = g^2~\int \frac{ d^2 k^{\prime}}{2\pi^2}
 Im \chi ({\bf k} - {\bf k}^{\prime}, \epsilon_k -
\epsilon_{k^{\prime}}) [n(\epsilon_{k^{\prime}} -\epsilon_k) +
f(\epsilon_{k^{\prime}})].
\end{equation}
where $n(\epsilon)$ and $f(\epsilon)$ are Bose and Fermi distribution
functions, respectively.
The integration over ${\bf k}^{\prime}$ can be split into an
integration over $\epsilon_{k^{\prime}}$ and an integration over surfaces
of
equal energy. For ${\bf k}$ located at the Fermi surface,
the first integral can be evaluated analytically.
  The exact result is rather cumbersome, but to a good numerical
accuracy it can be
approximated as
\begin{equation}
\frac{1}{\tau_k} =     {\alpha c^2_{sw}
g^2\over 8\gamma}\int_{FS} {dk^{\prime}\over |{\bf v}_F (k^{\prime})|}\,
{T^2\over \omega_{kk^{\prime}} (\omega_{kk^{\prime}}+\pi
T)}.
\label{eq:tau1}
\end{equation}
where $\omega_{kk^\prime}=\omega_{sf}(1 +
({\bf k-k^\prime-Q})^2\xi^2)$,
and ${k}$ and ${k}^\prime$ are two points at the Fermi surface.
The same expression is indeed obtained in an ordinary Fermi liquid. In
the latter case $\omega_{sf}$, and hence
$\omega_{kk^\prime}$ are of the order of the Fermi energy which is
large compared to $T$. One then immediately obtains $1/\tau_k \propto
T^2$ for
all ${k}$, i.e., the resistivity is proportional to $T^2$ as expected.
In cuprates, however,
$\omega_{sf}$ is much smaller than the Fermi energy;
the dominant contribution to the integral
over $k^{\prime}$ in (\ref{eq:tau1}) then
comes from the regions where the distance
between ${\bf k}^{\prime}$ (at the Fermi surface) and ${\bf k - Q}$
(at the ``shadow'' Fermi surface) is minimal. Clearly,
this minimal distance is zero at hot spots. Expanding
near each of the hot spots,
and performing the integration over $k^{\prime}$, we obtain
\begin{equation}
{\tau_{\bf k}} = \frac{4 |v^{hs}_F|}{\alpha \pi g^2 \xi
\sqrt{\omega_{sf}} }
\frac{Q_k}{T^2}
\label{eq:tau}
\end{equation}
where
\begin{eqnarray}
Q_k &=& \sqrt{\omega_{sf} (1+ (\Delta k)^2 \xi^2)}
\sqrt{\pi T+\omega_{sf}(1+(\Delta k)^2\xi^2)}\times \nonumber \\
&& \left(\sqrt{\omega_{sf}(1+
 (\Delta k)^2\xi^2)} + \sqrt{\pi T + \omega_{sf}(1+
(\Delta k)^2\xi^2)}\right),
\label{eq:Q}
\end{eqnarray}
Here $\Delta k$ is the displacement of ${\bf k}$ from
a nearby hot spot, and $v^{hs}_F$ is the Fermi velocity at the hot spots.
It is clear from Eq.\ (\ref{eq:Q}) that $\tau_k$ is the largest in
the regions of the Fermi surface where $\Delta k$ is greater than the inverse
correlation length, i.e., $(\Delta k)^2 \xi^2 \gg 1$.
These regions provide the dominant contribution to the conductivity,
while the regions near hot spots contribute very little.

On averaging $Q_k$ over the Fermi surface,
one finds that $<Q_k>$  is independent of
$T$ for $T <T_0$,
and $<Q_k>\sim T\sqrt{\omega_{sf}}\xi$ for $T>T_0$,
where the crossover temperature, $T_0$, is given by
\begin{equation}
T_0\approx \omega_{sf} (\Delta k_{max})^2\xi^2/2\pi.
\end{equation}
The extra factor of ${1\over 2}$ in $T_0$ arises for numerical reasons.
In general, one might expect that $T_0$ is of the order of the hopping
integral, i.e., it is much larger than $T$ over the whole
experimentally probed range of temperatures. If this was the case, then
the above reasoning would imply that the resistivity, $\rho\sim
T^2\xi\sqrt{\omega_{sf}}/<Q_k>$, would be
roughly quadratic in $T$. However,
photoemission experiments performed at or near
optimal doping show that a substantial
portion of the Fermi surface  is located near the magnetic Brillouin
zone boundary. It is not unreasonable to assume that this will be the
case for the overdoped materials as well.
In this case $T_0$ turns out to be much smaller
than the hopping integral. For example, in YBa$_2$Cu$_3$O$_7$\
on finds $T_0 \sim  15$meV$\ll t\sim
250$meV. In this situation $<Q_k>\sim T\sqrt{\omega_{sF}}\xi$ starting
already from relatively low temperatures and hence $\rho(T)\sim T$.
Our numerical analysis performed for the parameters of the
NAFLM chosen to fit NMR experiments in YBa$_2$Cu$_3$O$_7$
(Monthoux and Pines 1994a; Stojkovi\'c 1996) shows that
the resistivity is indeed roughly proportional to $T$ over
the temperature range $T\ge T_0\sim 15$meV.
Hlubina and Rice (1995) obtained $T_0\approx 20$meV with the same
values of spin-fluctuation parameters, but somewhat different band
parameters; hence they found $\rho\sim T^2$ up to higher
temperatures.  We emphasize once again
that the linear behavior of the
in-plane resistivity is due to the fact that $\omega_{sf}$
is the smallest parameter in the problem, even though it scales as $T$:
the typical $\omega_{sf} (\Delta k)^2\xi^2$ are smaller than $2\pi T$, but
much {\it larger} than $\omega_{sf}$.
If $\omega_{sf} \propto T$ was comparable to $\pi T$ above $T_{cr}$, then
the resistivity would be proportional to $T^{1/2}$.
Alternatively, if
$\omega_{sf}$ is large and nearly temperature independent,  as is the
case in the more heavily overdoped cuprates, then
 $\rho$ would be proportional to $T^2$, as seen experimentally.
%This, along with
% a particular choice of the band parameters, has led Hlubina and Rice
% (1995) to find $\rho\sim T^2$ at low temperatures even with reasonable
% choice of the spin fluctuation parameters.
Notice also that,  since $<Q_k>\propto T
\sqrt{\omega_{sf}}\xi$ above $T_{0}$,
the resistivity does not depend on the spin-damping and hence on
$\omega_{sf}\xi^2$. We  make use of
this fact below when we discuss the behavior of
resistivity at
smaller doping concentrations where $\gamma$ acquires a strong $T$
dependence.

It is also instructive to discuss the temperature dependence of the
average scattering rate, $\ll 1/\tau_k\gg$ for $T > T_{0}$.
Under the same conditions as above, $Q_k \propto T (\Delta k)$, and the
integration over $\Delta k$ yields $ <1/\tau_k> \propto T
\log ((\Delta k_{max})^2/\omega_{sf})$. However, the $T$ dependence
coming from the logarithm is rather weak and to a
good accuracy $<1/\tau_k> \propto T$. Therefore, we see that at $T >
T_{cr}$, inverting the average of $\tau_k$ and
the averaging of $1/\tau_k$ yields
a similar linear in $T$ dependence of the resistivity.

\section{Underdoped Cuprates}
\label{sec:underdoped}

We now turn to the key issue of the paper, the behavior of the
underdoped materials.  These may easily be distinguished
from overdoped materials on the basis of their low frequency
magnetic behavior, depicted in Fig.\ \ref{fig:6} (Barzykin and Pines 1995).
Examples of underdoped cuprates include YBa$_2$Cu$_4$O$_8$,
YBa$_2$Cu$_3$O$_{6+x}$, for $x\leq 0.93$, La$_{2-x}$Sr$_x$CuO$_4$ for
$x\leq 0.2$, Bi$_2$Sr$_2$CaCu$_2$O$_{8-\delta}$ (Bi 2212) compounds
with adjusted oxygen partial pressure during annealing and Hg 2223. As
seen in Fig.\ \ref{fig:6}, underdoped cuprates exhibit
two different crossovers in the normal state at the
temperatures $T_{cr}$ and $T_*<T_{cr}$. Above $T_{cr}$ their behavior
is similar to what has been observed for overdoped systems: it is
nonuniversal, but the resistivity is approximately linear in $T$, and
the ratio $^{63}T_1T/^{63}T_{2G}^2$ is independent of $T$, as shown for
YBa$_2$Cu$_4$O$_8$ in Fig.\ \ref{124nmr}a. The linear behavior of the
resistivity persists below $T_{cr}$ with only a subtle change in slope;
however, the magnetic properties change rather drastically below
$T_{cr}$. As shown in Fig.\ \ref{124nmr}b for YBa$_2$Cu$_4$O$_8$,
$^{63}T_1$
becomes almost independent of $T$, while $T_{2G}$ becomes proportional
to $T$ in such a way that the ratio $^{63}T_1T/^{63}T_{2G}$ is
independent of $T$. Using Eqs.\ (\ref{63Ta}) and (\ref{63Tb}), we find
that in this $T$ range the inverse correlation length and $\omega_{sf}$
both increase linearly with increasing $T$, in such a way that
$\omega_{sf}\xi=const$, implying $z=1$ behavior.
The uniform susceptibility
is also proportional to $T$ below $T_{cr}$. Finally, at even lower
temperatures, $T < T_*$, the
system enters into a regime where the correlation length becomes
independent of
$T$, $^{63}T_1 T \propto \omega_{sf}$ {\it increases} as $T$
decreases further towards
$T_c$ ($^{63}T_1 T$ displays a minimum at $T \sim T_*$), and
both the uniform susceptibility and $1/^{63}T_1T$ fall off sharply
with decreasing T.  This regime has been called a
``pseudogap regime" as
the behavior of, e.g., $\chi_0(T)$,
 is, at first sight, reminiscent of the behavior of systems which
display a true quasiparticle energy gap, such as a superconductor or an
ordered antiferromagnet. However, the curvature
of the fall-off in $\chi_0(T)$ is
opposite to that found below $T_c$ in conventional
superconductors (Slichter 1994).

We see therefore that the new features of the underdoped cuprates, as
illustrated in Figure \ref{fig:6}, are
($i$) the crossover from $z=2$ to $z=1$ behavior at $T_{cr}$, and ($ii$)
the crossover to pseudogap behavior at an even lower $T_*$.

Before presenting our scenario for these crossovers, we briefly review
the results of the
$\sigma-$model based studies of the
low-temperature crossovers between different {\it universal}
scaling regimes in disordered antiferromagnets
 (Chakravarty, Halperin and Nelson 1989,
Chubukov, Sachdev and Ye 1994, Sachdev, Chubukov and Sokol 1995).
 In the disordered state at $T=0$, the
antiferromagnets possess a gap, $\Delta$, in the excitation spectrum.
If the damping term at $T=0$ is much smaller than $\Delta$,
then,  as $T$ increases, the system experiences a crossover from a
quantum-disordered regime with an exponential ($e^{-\Delta/T}$) behavior of
observables, to the $z=1$ quantum-critical regime. If, on the contrary,
at
$T=0$, $\gamma \gg \Delta$, then the sequence of crossovers with
increasing $T$
is from a quantum-disordered regime with a {\it power-law} behavior of
observables
to the $z=2$ scaling regime with overdamped spin fluctuations, and
finally,
at even larger $T$, to the $z=1$ scaling regime in which  typical magnon
frequencies and the
damping term both scale linearly with $T$.

We see that for all ratios of $\gamma/\Delta$ the prediction of the
sigma-model based studies
is that $z=1$ behavior always occurs at higher temperatures than
$z=2$ behavior. This is simply due to the fact that  $z=1$ scaling
requires that
typical frequencies should be larger than the $T=0$ value of the damping
term.
Experimentally, however, the  situation is the opposite: in underdoped
cuprates,  one observes purely relaxational behavior at high enough
temperatures, and $z=1$ scaling at lower temperatures (Barzykin 1996).
The crossover at low $T$ from the quantum-disordered behavior to the
$z=1$ quantum-critical behavior is more consistent with the
observations around $T_*$, but again, the opposite curvature of the
fall-off in $\chi_0(T)$ compared to the sigma-model prediction indicates
that the crossover at $T_*$ involves not only localized spins, but also
electronic degrees of freedom.

Monthoux and Pines (1994b) proposed  that the reversed
behavior observed in underdoped cuprates below $T_{cr}$ is due to the
strong  temperature variation of
the  damping term $\gamma$
in the dynamical spin susceptibility, which in turn gives rise to the
anomalous temperature dependence of $\omega_{sf}$.
 This temperature variation is
neglected in the $\sigma$-model approaches which assume that throughout
the whole temperature range of interest,  the input parameters
in the dynamical susceptibility retain the same values as at $T=0$.
Below we relate
the temperature variation of the damping rate
 to the formation of a precursor to a spin-density-wave state which in
turn causes the
evolution of the quasiparticle Fermi surface with temperature
and doping concentration. We argue that
the two crossover scales $T_{cr}$ and $T_*$, observed in
underdoped cuprates indicate the
onset and the end point of this Fermi surface evolution, respectively.
At $T_{cr}$, the onset of the crossover,
the quasiparticle residue along the Fermi surface
develops a minimum at hot spots; at $T_*$, the end point of the
crossover,
the system actually begins to lose pieces of the Fermi surface.
In other words, we argue that
there exists only one extended crossover in system behavior in which
the electronic structure gradually
develops the features of a precursor to a spin-density-wave state.
In this scenario, $z=1$ scaling  in the temperature
range $T_*<T<T_{cr}$ is
just an intermediate asymptotic in the extended
crossover region, rather than an
extension of the  $z=1$ quantum-critical scaling observed in the
intermediate $T$ range right at half-filling.

 Fermi-surface evolution with doping at fixed $T$ has been observed
in photoemission experiments on YBCO (Liu et al 1992) and more
recently in  experiments on Bi-2212 (Marshall et al 1996,
LaRosa et al 1996). These experiments
demonstrated that while  near optimal doping the hole Fermi surface is
large and encloses an area consistent (to within the accuracy of
the measurement)
  with the Luttinger theorem, the measured Fermi surface in
 underdoped materials looses
 pieces near $(\pi,0)$ and symmetry-related points. Recent experiments
by Ding et al (1996) have shown that the same effect occurs at a fixed doping
as the temperature is lowered. The transformation
of spectral weight from
 the low frequency part of the spectrum
 to higher energies upon approaching half-filling
has been observed by Timusk and his collaborators in optical
experiments on  the planar conductivity
$\sigma(\omega,T)$ (Puchkov et al 1996); the same experiments also
demonstrated that there is no spectral weight transformation
at optimal (and larger) doping.

Recently, one of us has considered  Fermi-surface evolution
with increasing spin-fermion coupling
constant in the NAFLM at $T=0$ (Chubukov, Morr and Shakhnovich 1996).
It was found that as $g$ increases, the quasiparticle residue, $Z$
 near hot spots decreases. The
decrease of $Z$ becomes appreciable when the coupling constant $g$
exceeds
a critical value
$g^{(1)}_{cr} \sim v_F/(\omega_{sf} \chi_Q \log(C/\omega_{sf}))^{1/2}$.
Notice that $g^{(1)}_{cr}$ vanishes logarithmically  when the
correlation length becomes infinite. As $g$ increases even further, the
quasiparticle Fermi surface undergoes a substantial evolution in the
process of
which parts of the Fermi surface near the corners of the magnetic
Brillouin zone (where the hot spots are located when $g\rightarrow 0$)
move away and, simultaneously,  there appear two distinct
peaks in the density of states -
 the precursors of the valence and conduction bands. This evolution of
the
Fermi
surface occurs in the range of $g$ values
of the order of the upper cutoff in the staggered spin susceptibility,
$g \sim g^{(2)}_{cr} \sim C > g^{(1)}_{cr}$.
It was further argued
that as the system approaches
half-filling, the effective coupling constant $g$ increases, while
both $g^{(1)}_{cr}$ and $g^{(2)}_{cr}$  decrease,
 so that in varying the ratio $g/g_{cr}$ one in fact varies
the doping concentration.

Suppose now we fix the doping and vary the temperature.
As the temperature decreases, the inverse correlation length
and the  damping rate
decrease, and, hence,
both critical values of $g$ go down ($g^{(1)}_{cr}$ goes down chiefly because
of the decrease of damping, while $g^{(2)}_{cr}$ goes down chiefly because of
the decrease of $\xi^{-1}$). As a
result, even if the effective coupling  constant weakly depends on $T$,
the ratios, $g/g^{(1)}_{cr}$ and $g/g^{(2)}_{cr}$ still increase
 with decreasing
$T$. This in turn implies that, as the temperature is lowered, one
should observe
crossovers in the system behavior analogous to those described
above. These crossovers
will occur at temperatures at which $g^{(1,2)}_{cr} (T)$ become
 equal to a given $g$. It seems natural to associate the temperature
at which $g^{(1)}_{cr} (T_{cr})=g$ with $T_{cr}$,
and the temperature at which $g^{(2)}_{cr} (T_*)=g$ with $T_*$.

We now consider the conditions under which the system possesses
$z=1$ scaling behavior below $T_{cr}$.
Scaling with $z=1$ requires that typical frequencies and momenta be
of the  order
of $T$. In the NAFLM, this will be the case if
both $\xi^{-1}$ and $\omega_{sf}$
scale linearly with $T$, which in turn implies that
$\tilde{\Gamma}_{\bf {Q}_1}^{-1}$ and $\gamma$ should exhibit the same
linear dependence on T as does $\xi^{-1}$ (Chubukov, Sachdev and Ye 1994,
Monthoux and Pines 1994b).
Thus one should have between $T_{cr}$ and $T_*$,
\begin{equation}
\gamma \sim \tilde{\Gamma}_{\bf Q}^{-1} \sim \xi^{-1} = a+bT
\label{gammatilde}
\end{equation}
while, as we have seen, above $T_{cr}$, $\gamma$ and
$\tilde{\Gamma}_{\bf Q}^{-1}$ are independent of T.
As we discussed above,
the temperature-dependent inverse correlation length should be
considered as an
input parameter in the NAFLM because if one attempts to compute the real part
of the spin susceptibility (from which $\xi$ is inferred)
in RPA-type calculations, one would need to know the exact form of
the
fermionic Green's function far away from the Fermi surface.
Hence, the linear in $T$ dependence of
$\xi^{-1}$ below $T_{cr}$  follows from the fit to the NMR
data on $^{63}T_1$ and $T_2$ (Barzykin and Pines 1995).
On the contrary,  the linear $T$ dependence of $\gamma$ has to be
obtained
 {\it within} the NAFLM approach, although at the moment we do not have a
clear recipe for calculating
it. 

It is essential  however not to confuse the intermediate
$z=1$ scaling regime in underdoped cuprates with the $z=1$
quantum-critical
scaling in pure antiferromagnets. In the latter case,
the damping of spin excitations is due to the interaction between spin
fluctuations, and one can argue quite generally that it should be linear
in $T$ when $\xi^{-1} \propto T$ (Chubukov, Sachdev and Ye 1994).
 However, the estimates  of
Barzykin and Pines (1995)
show that at doping concentrations where experiments
on underdoped cuprates have been performed,
the purely magnetic damping is
too small to account for the experimental data. In other words, in the
$z=1$ regime in underdoped cuprates, the damping primarily comes from
the interactions with fermions (this is why we describe this regime as
``pseudoscaling''). At the same time we expect that below $T_{cr}$,
lattice
corrections are not large. Then, once the $z=1$ form of the dynamical
susceptibility for ${\bf q}$ near ${\bf Q}$
is obtained, one can use the universal scaling forms
obtained for $z=1$ quantum-critical scaling for both uniform and
staggered susceptibility. In particular, this implies that
the static uniform susceptibility should be linear in $T$, in agreement with
experiment.
This, we recall, is different from the $z=2$ regime at higher
temperatures,
where the physics is dominated by short wavelength fluctuations and
one could only formally use the predictions of the $z=2$
quantum-critical
scaling to obtain the $T$ dependence of observables associated with the
dynamical susceptibility near ${\bf Q}$; however, the temperature
dependence
of the
uniform susceptibility was  very different from the scaling prediction.

We note in passing that the fact that spin damping is chiefly due to
the interaction with fermions may resolve
an apparent discrepancy between the values of the spin-wave
velocity in $YBa_2Cu_3O_{6.63}$
 extracted from a linear fit to the uniform susceptibility  and from the
ratio of $^{63}T_1T/T_{2G}$ (Millis and  Monien 1994,
Chubukov, Sachdev and Sokol 1994). Indeed, the slope of the uniform
susceptibility $\chi_0 \propto T$ is finite in the absence of spin damping.
It does not change much with doping compared to its value at half-filling,
and
therefore is likely to only weakly depend on $\gamma$.
Accordingly, the fit for $\chi_0 (T)$ yields a
$c_{sw}$ which is
slightly larger than that found at half-filling, which  agrees with
analytical calculations at small doping
(Chubukov and Frenkel 1992, Schulz and Zhou 1995).
At the same time, the NMR relaxation rate is proportional to the
damping, and
for this quantity it is essential to know
where the damping comes from.
A fit to NMR data using the $\sigma$-model results yields
a spin-wave velocity  which is a few times smaller than at
half-filling. If instead, we
extract $c_{sw}$ from the NMR data using Fermi-liquid damping,
we obtain a spin-wave velocity larger by a factor
$\gamma_{FL}/\gamma_{sf}$ which better agrees with the velocity
extracted from the  susceptibility data.

We consider next the behavior of the resistivity.
Experimentally, it is linear in $T$
both above and below $T_{cr}$ with about the same slope.
On the theoretical side, we have seen above that the
resistivity is linear in $T$ as long as over substantial portions of the
Fermi surface, $T$ is larger than $T_0$.
This result holds approximately
even when $\gamma$ acquires a substantial $T$
dependence, provided the vertex corrections are not too large.
More importantly,
the decrease of the spectral weight near the hot spots has little effect
on the conductivity, simply because the dominant contributions to
conductivity come from
the pieces of the Fermi surface which are relatively far from hot spots.
A careful inspection of the measured  shape of the Fermi surface in YBCO
compounds (Gofron et al 1994, Campuzano et al 1994)
and the fits to NMR measurements (Barzykin and Pines 1995),
shows that in the underdoped cuprates
$T_0$ is also substantially smaller than $T_{cr}$. In the
intermediate, $z=1$ scaling regime, $\gamma$ acquires a temperature
dependence, and $T_0$ becomes a function of $T$, scaling
as $T_0\sim T^{-1}$. However, the slope is rather small and everywhere
in the $z=1$ regime $T_0$ remains smaller than $T$, i.e., the
resistivity remains linear in $T$, in good agreement  with the
experimental results in, e.g., YBa$_2$Cu$_4$O$_8$ (Bucher et al 1993,
see Fig.\ \ref{124rho}). The two temperature scales ($T$ and $T_0(T)$)
become comparable to each other at a temperature which in
YBa$_2$Cu$_4$O$_8$ is rather close to $T_*$ ($T_*$ is roughly 220K and
$T_0(T_*)\sim250$K). We emphasize, however, that the behavior of
resistivity in our model is nontrivial and depends on the shape of the
quasiparticle band structure. Hence our argument, based on
YBa$_2$Cu$_4$O$_8$, that the crossover in resistivity occurs near $T_*$
need not necessarily apply to all high-T$_c$ compounds.

We now turn to the crossover at $T=T_*$.  Barzykin and Pines (1995)
associated this
crossover with the transformation from
the $z=1$ quantum-critical to the $z=1$ quantum-disordered regime.
We have associated the same crossover
with the development of the precursors of the spin-density-wave state in
the
electronic structure.
We now show that the two identifications of the crossover
at $T_*$ are in fact complementary.
We note first that in order to obtain
predominantly $z=1$  behavior for the observables
in the quantum-disordered regime, one must
get rid of the Landau damping term in
the dynamical staggered susceptibility.
Otherwise, the system at the lowest $T$
will necessarily crossover to the quantum-disordered
$z=2$ regime, where, e.g., $^{63}T_1 T \propto \omega_{sf} = const$,
which does not agree with the data below $T_*$ (we recall that the measured
$^{63}T_1T$ {\it increases} as the temperature is lowered below $T_*$).
What happens when the  quasiparticle
spectrum develops a pseudogap near the corners of the Brillouin zone?
At first glance, one might argue that the spin
 damping must
become exponentially small in $T$ at low temperatures
because the former hot spots which chiefly contributed to the damping in
the
overdoped regime disappear.
However, this argument is incorrect
because in a process of the Fermi surface evolution towards
small pockets, there appear new hot spots which survive
even in a situation in which the spin-density-wave structure of
the electronic states is already developed.

The actual reason the damping term goes down sharply 
when pockets are formed relates to the form of the fully
renormalized spin-fermion vertex: in the spin-density-wave state with
long-range order this vertex vanishes
identically for the bosonic momentum $Q$ because of
the Ward identity (Adler principle); thus,
the damping term in the spin susceptibility in fact scales as
$Im \chi^{-1} (q,\omega) \propto i \omega (q-Q)^2$ (Sachdev 1994,
Schrieffer 1995, Sachdev, Chubukov and Sokol 1995). In the precursor to
the spin-density-wave state, there is
no precise requirement that the full vertex at $Q$ should vanish;
however explicit calculations show that vertex corrections (which
in the $g \gg
C$ limit  have a negative sign) almost completely cancel the bare
interaction $g$ such that
the fully renormalized vertex turns out to be
 small compared to $g$ by a factor $(g^{(2)}_{cr}/g)^2$
(Chubukov, Morr and Shakhnovich 1996). Very similar arguments have been
previously provided by Schrieffer (Schrieffer 1995).

A nice feature of the large $g$
limit in the spin-fermion model is that the smallness of
$g^{(2)}_{cr}/g$ not only yields a small vertex but also allows one to
compute the full vertex and full fermionic Green's function by
expanding in powers of $g^{(2)}_{cr}/g$. This in turn allows one to
compute fermionic damping: one just substitutes the full
Green's functions and full vertex into the particle-hole bubble.
We have performed this calculation and found that the damping term
contains a small factor $(g^{(2)}_{cr}/g)^4$.
Since the ratio $ g/g^{(2)}_{cr}$ decreases with $T$,
the damping term rapidly decreases with decreasing
temperature, and at low temperatures becomes smaller than $c_{sw}
\xi^{-1}$
which is the energy scale of a gap in the spin susceptibility.  
In this situation,
the system should display predominantly $z=1$ quantum-disordered behavior,
in
agreement with the data below $T_*$. In particular, $\omega_{sf}
\propto 1/\gamma$
should {\it increase} as $T$ decreases which in turn leads to the
{\it increase}
of $^{63}T_1 T$ with decreasing temperature.
In addition, below $T_*$, the resistivity scales as
$\rho \propto T^2 /\omega_{sf}$, i.e., it rapidly (faster than $T^2$)
decreases with decreasing
$T$. This rapid decrease of $\rho$
is also consistent with the  experimentally observed
behavior of the resistivity (see Fig {\ref{124rho})).
Finally, at even lower temperatures,  the system should, in
principle,
undergo a crossover to $z=2$ quantum-disordered regime, but the
crossover temperature is likely lower than $T_c$ which implies that
this crossover cannot be observed experimentally.

The reduction of the spin-fermion interaction vertex
also affects the low $T$ behavior of the uniform susceptibility. In a
fully developed spin-density-wave
state with long-range order only the longitudinal
susceptibility has a (doping dependent) Pauli contribution, associated
with a finite density of holes; the transverse susceptibility
$\chi_\perp$ at finite
doping remains virtually the same as at half-filling because the Pauli
contribution to $\chi_\perp$
is reduced by vertex renormalization (Chubukov and Frenkel
1992, Chubukov and Musaelian 1995). Specifically, we have
\begin{equation}
\chi_\perp^{(T=0)} = {Z_\chi\over 4J} + O(x);\qquad \chi_{zz}^{(T=0)} =
{1\over 2} \chi_\rho
\end{equation}
where $x$ is the doping concentration, $Z_\chi$ is the quantum
renormalization factor, and for a Fermi surface consisting
of small pockets
near $(\pi/2,\pi/2)$, one finds $\chi_\rho=\sqrt{m_1 m_2}/2\pi$, where
$m_1$ and $m_2$ are the two effective masses for fermionic dispersion.
In order to estimate $\chi_\rho$, we use the photoemission data for the
oxychloride Sr$_2$CuO$_2$Cl$_2$ (Wells et al 1995, LaRosa et al 1996). 
These data
yields $m_1\approx m_2\approx 1/2J$, or $\chi_\rho\approx 1/4\pi J$. In
the preformed spin-density-wave
state the transverse and longitudinal
susceptibilities are indistinguishable so that
\begin{equation}
\chi_u (T\rightarrow 0)\equiv {1\over 3} \chi_\perp + {1\over 3}
\chi_{zz} = {Z_\chi\over 12 J} + {1\over 6} \chi_\rho.
\end{equation}

Consider further the underdoped cuprates with a magnetically disordered
ground state and a small Fermi surface.
As the correlation length saturates to a finite value as
$T\rightarrow 0$, the magnetic part of the susceptibility vanishes
($Z_\chi\rightarrow 0$). We are thus left with only the Pauli contribution,
i.e., $\chi_u=(1/3) (\chi_\rho/2)\approx 1/24\pi J$. We emphasize that
the factor 1/3 in $\chi_u$ is due to the spin-fermion vertex reduction,
i.e., the same effect which yields a reduction in the spin damping.
Reinserting the $(g\mu_B)^2$ factor in $\chi_u$ we obtain
$\chi_u/\mu_B^2\sim 0.4$states/eV per Cu atom which is some six times
smaller than $\chi_u/\mu_B^2\geq 2.6$states/eV, the result for an
optimally doped sample such as YBa$_2$Cu$_3$O$_7$, or for overdoped
samples. We see therefore that
the uniform susceptibility in the preformed spin-density-wave
state with a small
Fermi surface is much smaller than that found
at larger doping levels with large
Fermi surfaces, despite the fact that the Pauli susceptibility in 2D
does not depend on $p_F$ and is the same for large and small elliptical
Fermi surfaces.

\section{Optimal Doping}
\label{sec:optimal}

Optimally doped cuprates are frequently defined as those which,
within a given family,
exhibit the highest $T_c$. Examples of optimally doped materials
include, e.g., YBa$_2$Cu$_3$O$_{6.93}$ and
La$_{1.85}$Sr$_{0.15}$CuO$_4$. Photoemission experiments on YBCO
systems and the bismuthates have shown that these materials have a
large Fermi surface in close analogy with overdoped cuprates. At the
same time, from a magnetic perspective, optimally doped materials are
members of the underdoped family, in that their uniform susceptiblity
exhibits the same crossovers at $T_{cr}$ and $T_*$ as are seen in other
underdoped materials. The analogy between underdoped and optimally
doped materials has also been observed in photoemission experiments on
Bi 2212 (Ding et al 1996): in both types of materials, there exists a
pseudogap in the normal state which disappears at $T\sim T_{cr}$ which
is substantially larger than $T_c$.

Another feature of optimally doped materials is that their resistivity
continues to be linear in $T$ down to $T\sim T_c$. According to the
arguments we have given above, such linearity above $T_0$ reflects both
the closeness of the Fermi surface to the magnetic Brillouin zone
boundary and the particular spin-fluctuation spectrum. It is tempting
to conjecture that in the optimally doped cuprates the Fermi surface is
rather flat in such a way that $(\Delta k_{max})^2$ is minimized, which
in turn yields a minimum in $T_0$ as a function of doping.
This flattening of the Fermi surface would also give
rise to stronger vertex
corrections which, as we recall, act to increase
the spin-fermion interaction
vertex and hence $T_c$, before any changes in the Fermi surface
topology take place.

\section{Summary}
\label{sec:conclusion}

We have proposed a specific  scenario for the
temperature crossovers in the overdoped and  underdoped cuprates.
We considered
 a nearly
antiferromagnetic Fermi-liquid model and argued
 that in the overdoped regime the spin damping, $\gamma\propto
(\omega_{sf}\xi^2)^{-1}$ is independent of temperature over the entire
experimentally probed $T$-range. In this situation the magnetic
behavior is described by a mean-field $z=2$ dynamical susceptibility.
We argued that the resistivity has a crossover from a Fermi-liquid like
$T^2$ behavior at $T<T_0$ to a linear in $T$ behavior for $T>T_0$; the
crossover temperature $T_0$ is low due to the proximity of the Fermi
surface to the magnetic Brillouin zone boundary.
We further argued that in order
to account for the experimentally measured sequence of crossovers in
underdoped
cuprates from
the Fermi-liquid, $z=2$
 behavior at high temperatures to the $z=1$ scaling behavior
at intermediate $T$, and to the  pseudogap behavior at even lower
$T$, one
has to take into consideration the thermal variation of the damping rate
of spin excitations, $\gamma$.
We argued that the
primary source of the variation of $\gamma$ with $T$ in the underdoped
cuprates is the thermal evolution of the quasiparticle dispersion
near the Fermi surface produced by the precursor of a
spin-density-wave state. Within our approach, we found
consistency with the experimental data on resistivity, NMR relaxation rates
and uniform susceptibility in underdoped
cuprates (for a discussion of the application of NAFLM to the Hall
effect measurements see Stojkovi\'c 1996).
We emphasize, however, that here
we only discuss a possible scenario of how the damping evolves with $T$. At
present, we can compute this damping either for very small or very large
values of the coupling constant.  Direct  calculations of
$\gamma (T)$ within the spin-fermion model at intermediate couplings are
clearly called for.

A final note.  Our scenario is in contradiction with the proposals
(Emery and Kivelson 1995, Randeria et al 1994, Ding et al 1996) that the
 physical origin of the pseudogap behavior
is the precursor to the $d-$wave pairing state.
%  In this last scenario,
%  the superconducting gap appears above $T_c$ due to superconducting
%  fluctuations.
Both this and our scenario imply that the quasiparticle gap
near $(0,\pi)$ observed in photoemission measurements in the normal state
of underdoped cuprates should not change as the system becomes
superconducting, in agreement with the data.
In the precursor to the $d-$wave pairing scenario,
the superconducting gap is already
preformed by superconducting fluctuations, while in our scenario there
exists a preformed spin-density-wave gap near $(0,\pi)$.

The photoemission data of Ding et al (1996) have been interpreted as an
evidence in favor of preformed d-wave gap.
They measured a gap as a shift of the midpoint of the leading edge of
the photoemission spectrum, and found that the shift, which is of the
order of 20meV, has the same $k$-dependence as the d-wave gap in the
superconducting state. On the contrary, Marshall et al (1996)
estimated  the gap from the position of the maximum in the
spectral function. They found a 
much larger gap of about $0.2 eV$ for  $60K$ superconductor, which  
is fully consistent with spin-density-wave scenario
(in a fully developed spin-density-wave state, the
gap near $(0,\pi)$ is about $2J\sim 0.25$eV). In the
absence of a theory for the lineshape, it is difficult to say with
confidence which interpretation of the data is correct.
We note, however, that no preformed gap has been observed in overdoped
cuprates. While the magnetic scenario leads naturally to the
prediction that there should be no sign of a pseudogap in overdoped cuprates
(superconductivity occurs before spin-density-wave precursor can develop),
it is not clear why precursors to the $d-$wave pairing state should disappear
in overdoped materials.
Finally, we note that the magnetic scenario  correctly describes
the whole sequence of crossovers in the normal state including the
crossover at $T_{cr}$ which can easily be much larger than $T_c$
(e.g., $T_{cr}\approx 6 T_c$ in YBa$_2$Cu$_4$O$_8$).
Moreover, NMR experiments on YBa$_2$Cu$_3$O$_7$, YBa$_2$Cu$_4$O$_8$ and
La$_{1.85}$Sr$_{0.15}$CuO$_4$ suggest that the magnetic correlation
length is appoximately the same at $T_{cr}$,  $\xi(T_{cr})\sim 2a$,
in all of these compounds, implying yet another
connection between the magnetic and the pseudogap behavior.
It also seems unlikely that one could explain the crossover to $z=1$
scaling as due to a precursor of the d-wave pairing.
We therefore
believe that the pseudogap has a magnetic rather than superconducting
origin.

Several researchers (see, e.g., Millis and Monien 1994) have argued
that the pseudogap behavior in the underdoped cuprates is due to an
exchange coupling, $J_\perp$, between the bilayers found in a unit
cell. Indeed, strong bilayer coupling leads to singlet configurations
of adjacent spins which gives rise to a gap in the spin excitation spectrum.
In an insulator, however, one needs a rather large $J_\perp$ ($\sim 2.J$)
to produce the single configuration between adjacent spins in the
bilayers, while recent experiments find $J_\perp\sim 0.1 J$ (Keimer et
al 1996). For doped materials, Ioffe et al (1994) argued that
the effective bilayer coupling scales as $J^{eff}_{\perp} \propto J_{\perp}
\chi^2(Q)$, i.e., it is enhanced if susceptibility is strongly peaked at $Q$.
This effect is certainly present at optimal doping, however we argued above
that in the underdoped cuprates (where pseudogap has been observed), the
enhancement of susceptibility is compensated by vertex corrections. In this
sutuation, $J^{eff}$ should be of the same order as the bare coupling and is
unlikely to give rise to a spin gap unless one assumes that there is a
spin-charge separation
(Millis, Altshuler and Ioffe 1996, Ubbens and Lee 1994).
Our point of view is that  the
 pseudogap behavior is an intrinsic property of a single
CuO$_2$ layer as  the data
on the  spin-lattice relaxation rate, the uniform
susceptibility, and $\xi(T)$, in the single layer
La$_{2-x}$Sr$_x$CuO$_4$ materials display the same sequence of
crossovers seen in the bilayer YBa$_2$Cu$_3$O$_{6+x}$ materials.

It is our pleasure to thank A. Millis for helpful conversations and a careful
reading of the manuscript. We thank V. Barzykin, B.\ Batlogg,
G.\ Blumberg,
H. Monien,  M. Onellion, S. Sachdev, D. Scalapino, C. Slichter, R.
Schrieffer, Z-X Shen, A. Sokol, R. Stern  and T. Timusk
for numerous discussions and comments. We also thank N. Curro for
providing us with the NMR data on 124 compound prior to publication.
A.C. and D.P. acknowledge the hospitality of
ITP, Santa Barbara, where part of this work has been performed.
The research at ITP has been supported in part by a NSF grant
PHY94-071194.
A.C.\ is an A.P.\ Sloan fellow. D.P.\ and B.P.S.\ are sponsored in part by
NSF grants NSF-DMR 89-20538 (Materials Research Laboratory at the
University if Illinois at Urbana-Champaign) and NSF-DMR 91-20000 (Science
and Technology Center for Superconductivity).

\vfill\eject

\section*{References}

\begin{itemize}
%\begin{harvard} %J. Phys: Cond. Mat.

\item %{altshuler}
B. L. Altshuler, L. B. Ioffe and A. J. Millis,
Phys.\ Rev.\ B {\bf 52}, 415 (1995).

\item %{stamp}
M. H. S. Amin and P. C. E. Stamp, preprint (1996).

\item %{pwa}
P. W. Anderson, Science {\bf 235} 1196 (1987).

\item %{pwa-photo}
P. W. Anderson, Rev. Math. Phys. {\bf 6}, 1085 (1994).

\item %{bpst}
V. Barzykin, A. Sokol, D. Pines and D. Thelen, Phys.\ Rev.\ B {\bf 48}, 1544
(1993).

\item %{bp}
V. Barzykin and D. Pines, Phys.\ Rev.\ B {\bf 52}, 13585 (1995).

\item %{victor}
V. Barzykin, to appear in Phil.\ Mag., (1996).

\item %{basov}
D. N. Basov, R. Liang, B. Dabrowski, D. A. Bonn, W. N. Hardy,
and T. Timusk, preprint (1996).

\item %{ybco-ins}
P. Bourges, L. P. Regnault, Y. Sidis, C. Vettier, Phys.\ Rev.\ B {\bf 53}
876 (1996).

\item %{bucher}
B.\ Bucher, P.\ Steiner, J.\ Karpinski, E.\ Kaldis and P. Wachter, 
Phys.\ Rev.\ Lett.\ {\bf 70}, 2012 (1993).

\item %{bulut}
N. Bulut, D.J. Scalapino and S.R. White, Phys.\ Rev.\ B {\bf 47}, 2742 (1993).

\item %{camp}
J. C. Campuzano, K. Gofron, H. Ding, R. Liu,
B. Dabrowski, B. Veal, J. Low Temp. Phys. {\bf 95}, 245 (1994).

\item %{chn}
S. Chakravarty, B. I. Halperin and D.
Nelson, Phys.\ Rev.\ Lett.\ {\bf 60}, 1057 (1988).

\item %{pwa-chak}
S.\ Chakravarty and P.\ W.\ Anderson, Phys.\ Rev.\ Lett.\
{\bf 72}, P3859 (1994).

\item %{chufre}
A. V. Chubukov and D.\ Frenkel, Phys.\ Rev.\ B {\bf 46}, 11884 (1992).

\item %{csy}
A. V. Chubukov, S. Sachdev and J. Ye, Phys.\ Rev.\ B {\bf 49}, 11919 (1994).

\item %{css}
A. V. Chubukov, S. Sachdev and A. Sokol, Phys.\ Rev.\ B {\bf 49}, 9052 (1994).

\item %{gammafl}
A. V. Chubukov, Phys.\ Rev.\ B {\bf 52}, R3840 (1995).

\item %{fsevol}
 A. V. Chubukov , D. K. Morr , K. A. Shakhnovich, to appear in Phil.\
Mag.\ 1996.

\item %{curro}
N. Curro, R. Corey, C. P. Slichter, preprint, (1996).

\item %{ding}
H.\ Ding et al, Nature {\bf 382}, 51 (1996).

\item %{emery}
V. K. Emery and S. A. Kivelson, Phys.\ Rev.\ Lett.\ {\bf 74}, 3253 (1995).

\item %{gofron}
K. Gofron  et al, Phys. Rev. Lett {\bf 73}, 3302 (1994).

\item %{gruner}
G. Gr\"uner, {\em Density Waves in Solids}, (Addison-Wesley, Reading,
1994).

\item %{hayden}
F. M. Hayden et al, Phys.\ Rev.\ Lett.\ {\bf 76}, 1344 (1996).

\item %{tmrice}
R.\ Hlubina and T.\ M.\ Rice,
Phys. Rev. B 51, 9253 (1995); {\em ibid} 52, 13043 (1995).

\item %{hwang}
H. Y. Hwang et al, Phys.\ Rev.\ Lett.\ {\bf 72}, 2636 (1994).

\item
L. B. Ioffe, A. I. Larkin, B. L. Altshuler, and A. J. Millis,
JETP Lett. {\bf 59}, 65 (1994).

\item %{itoh}
Y. Itoh et al, J. Phys. Soc. Japan 63, 22 (1994).

\item %{iye}
Y. Iye, in {\em Physical Properties of High Temperature
Superconductors}, ed.\ by D.\ M.\ Ginsberg (World Scientific, Singapore,
1992, Vol.\ 3.

\item %{johnson}
D. C. Johnson, Phys.\ Rev.\ Lett.\ {\bf 32} 957 (1989).

\item %{meinkampf}
A. P. Kampf, Phys. Rep. {\bf 249}, 219 (1994).

\item %{keimer}
B.\ Keimer et al, preprint (1996).

\item %{kita}
Y. Kitaoka et al, Physica C 179, 107 (1991).

\item
S. LaRosa et al, preprint (1996).

\item %{lee}
P. A. Lee and N. Nagaosa, Phys.\ Rev.\ B {\bf 46}, 5621 (1992);
N. Nagaosa and P. A. Lee, Phys. Rev. B {\bf 45}, 966 (1992).

\item
R. Liu et al, Phys.\ Rev.\ B {\bf 46}, 11065 (1992).

\item %{marshal}
D. S. Marshall et al, Phys.\ Rev.\ Lett. {\bf 76}, 4841 (1996);

\item %{aeppli}
T.\ E.\ Mason, G.\ Aeppli, S.\ M.\ Hayden, A.\ P.\ Ramirez
and H.\ A.\ Mook, Physica B {\bf 199}, 284 (1994).

\item %{mila-rice}
F. Mila and T. M. Rice, Physica C 157, 561 (1989).

\item
A. J. Millis, B. L. Altshuler and L. B. Ioffe,
Phys.\ Rev.\ B {\bf 53}, 415 (1996).

\item %{MMP}
A.\ Millis, H.\ Monien and D.\ Pines, Phys.\ Rev.\ B {\bf 42}, 167 (1990).

\item%{andy}
A. Millis, Phys.\ Rev.\ B {\bf 48}, 7193 (1993).
\item %{milmon}
A. J. Millis and H. Monien, Phys.\ Rev.\ B {\bf 50}, 16606 (1994).

\item %{MP}
P.\ Monthoux and D.\ Pines, Phys.\ Rev.\ B {\bf 47}, 6069 (1993).

\item %{MP}
P.\ Monthoux and D.\ Pines, Phys.\ Rev.\ B {\bf 50}, 16015 (1994a).

\item %{MP}
P.\ Monthoux and D.\ Pines, Phys.\ Rev.\ B {\bf 49}, 4261 (1994b).

\item %{private}
P.\ Monthoux, unpublished, (1996).

\item %{ong}
N.\ P.\ Ong in {\em Physical Properties of High Temperature
Superconductors}, ed.\ by D.\ M.\ Ginsberg (World Scientific, Singapore,
1990), Vol.\ 2.

\item %{david95}
D.\ Pines, in {\em High Temperature Superconductivity and
the C$^{60}$ Family},
ed.\ H.\ C.\ Ren, p.\ 1 (Gordon and Breach, 1995).

\item %{puchkov}
A. V. Puchkov, P. Fournier, T. Timusk and N. N. Kolesnikov, preprint
(1996).

\item %{randeria}
M.\ Randeria, N.\ Trivedi, A.\ Moreo and R.\ T.\ Scalettar, Phys.\
Rev.\ Lett.\ {\bf 72},  3292 (1994).

\item %{rickayzen}
G.\ Rickayzen, in {\em Green's Functions and
Condensed Matter Physics}, (Academic Press, London, 1980).

\item %{sach94}
S. Sachdev, Phys.\ Rev.\ B {\bf 49}, 6770 (1994).

\item %{scs}
S. Sachdev, A. V. Chubukov and  A. Sokol, Phys.\ Rev.\ B {\bf 51},
14874-14891 (1995).

\item %{scalapino_review}
D.J. Scalapino, in {\it ``Proceedings of the international school of physics
Enrico Fermi"}, ed.\ by R.A. Broglia and J.R. Schrieffer, North-Holland (1994)
and references therein.

\item %{schrieffer}
J. R. Schrieffer, X. G. Wen, and S. C. Zhang,
Phys.\ Rev.\ B {\bf 39}, 11663 (1989).

\item %{bob}
J.\ R.\ Schrieffer, J.\ Low Temp.\ Phys.\ {\bf 99}, 397 (1995).

\item %{schultz}
H.J. Schulz and  C. Zhou, Phys.\ Rev.\ B {\bf 52}, 11557 (1995).

\item %{shastry}
B. Shastry, Phys.\ Rev.\ Lett.\ {\bf 63}, 1288 (1989).

\item %{shen}
Z.-X. Shen et al, unpublished (1996).

\item %{si}
Q. Si, Y. Zha, K. Levin and J.P. Lu, Phys. Rev. B {\bf 47}, 9055
(1993).

\item %{charlie}
C,\ P.\ Slichter, in {\em Strongly Correlated Electronic
Systems}, ed.\ by K.\ S.\ Bedell {\em et al}, (Addison Wesley, 1994).

\item %{sokol-pines}
A. Sokol and D. Pines, Phys. Rev. Lett. {\bf 71} 2813 (1993).

\item %{shraiman}
B.\ I.\ Shraiman and E.\ D.\ Siggia, Phys. Rev. Lett. {\bf 61}, 467 (1988).

\item %{bps}
B. P. Stojkovi\'c, to appear in Phil.\ Mag. (1996).

\item %{thelen}
D.\ Thelen and D.\ Pines, Phys.\ Rev.\ B {\bf 49}, 3528 (1994).

\item
M. U. Ubbens and P. A. Lee,  Phys.\ Rev.\ B {\bf 50}, 438 (1994).

\item %{varma}
C. Varma et al, Phys. Rev. Lett. {\bf 63}, 1996 (1989).

\item %{wells}
B. O. Wells et al, Phys.\ Rev.\ Lett.\ {\bf 74} 964 (1995).

\item %{zbp}
Y. Zha, V. Barzykin and D. Pines, to appear in Phys.\ Rev.\ B. (1996).

\end{itemize}

% \end{harvard}  %J. Phys: Cond. Mat.

\vfill\eject

\begin{figure}
\caption{The schematic behavior of various
observables in underdoped cuprates (after Barzykin and Pines 1995). 
The temperature scales
$T_{cr}$ and $T_*$ indicate crossovers which are defined and discussed
in the text.}
\label{fig:6}
\end{figure}

\begin{figure}
\caption{(a) Copper spin-lattice
relaxation rate $^{63}T_1T$ as a function of temperature
in slightly overdoped Tl$_2$Ba$_2$CuO$_{6+x}$ (Itoh et al 1994).
Above $150$K
$^{63}T_1T$ and, hence, $\omega_{sf}$ is proportional to $T$. At lower
temperatures,
$^{63}T_1T$ becomes $T$ independent as in a conventional Fermi-liquid; (b) the
ratio
$^{63}T_1T/T^2_{2G}$, where $T_{2G}$ is the spin-echo decay time.
This ratio is practically
independent of temperature, which is consistent with $z=2$ scaling.}
\label{nmrover}
\end{figure}

\begin{figure}
\caption{A sequence of crossovers in  overdoped (a) and
underdoped (b) cuprates.}
\label{diagram}
\end{figure}

\begin{figure}
\caption{The resistivity as a function of temperature for two overdoped
Tl$_2$Ba$_2$CuO$_{6+\delta}$ samples (Kitaoka et al 1991)
with $T_c=40$K (upper curve)
and $T_c=0$K (lower curve). The second sample is more heavily
overdoped.
Notice a pronounced $T^2$ behavior in both samples at very low
temperatures. The top curve clearly exhibits a
crossover to $\rho\propto T$ somewhere between $100$ and $200$K. The
crossover
temperature for the bottom curve is presumably larger than $300K$.}
\label{rhoover}
\end{figure}

\begin{figure}
\caption{(a) The ratio $^{63}T_1T/T^2_{2G}$ as a function of temperature
 in underdoped
YBa$_2$Cu$_4$O$_8$. As in overdoped samples,
this quantity becomes  independent of temperature above $T_{cr}\sim
450-500$K;
(b) $^{63}T_1T/T_{2G}$ as a function of temperature for the same compound.
This ratio is almost a constant in
the temperature interval $T_* < T < T_{cr}$, which  is a
 signature of the $z=1$ scaling regime. In both graphs,
 the dashed line is a guide to the eye.}
\label{124nmr}
\end{figure}

\begin{figure}
\caption{The resistivity along the $a-$axis  for underdoped
YBa$_2$Cu$_4$O$_8$.
Notice that the linear behavior starts at
around $T_*$, and 
the minor change of slope at $T=T_{cr}$. Below $T_*$, the
resistivity rapidly falls off with decreasing $T$.}
\label{124rho}
\end{figure}

\vfill\eject
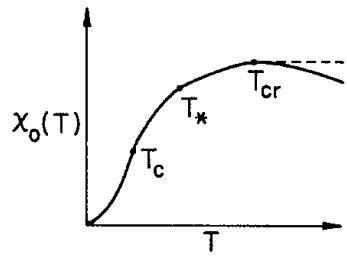
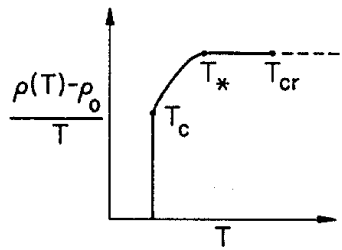
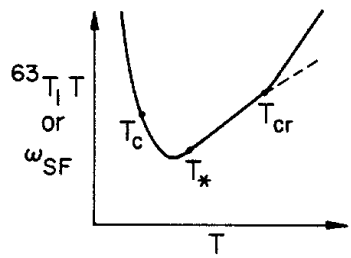
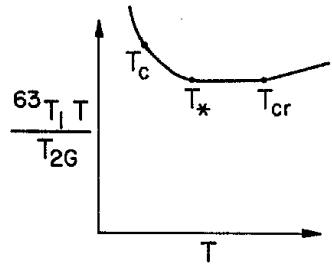
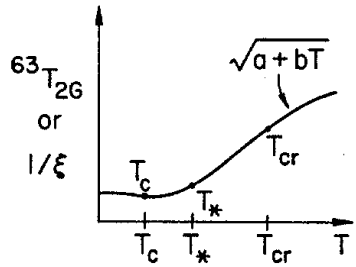
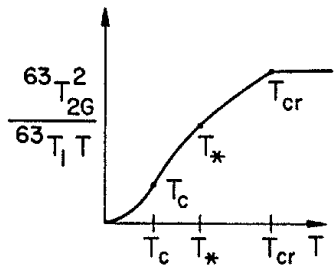
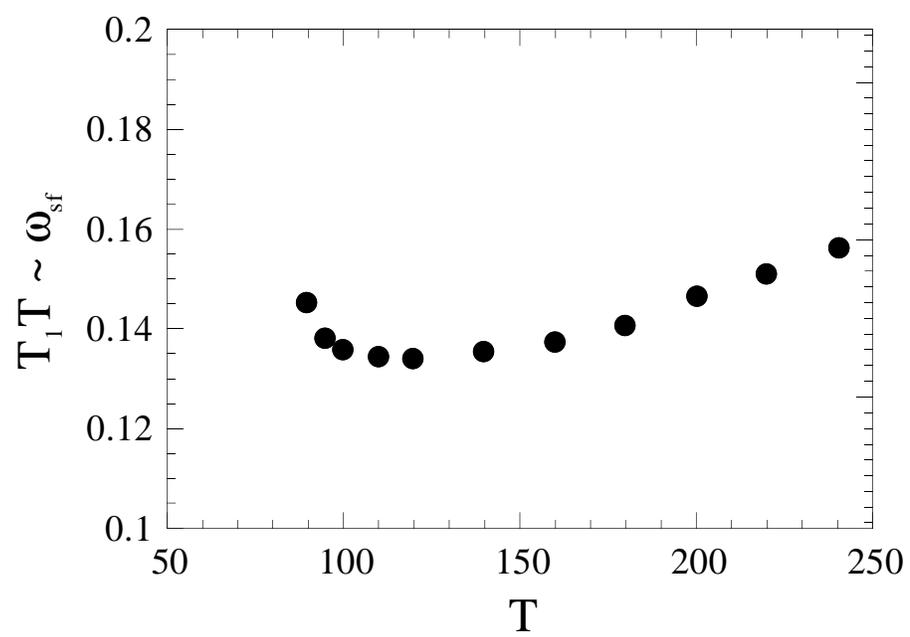
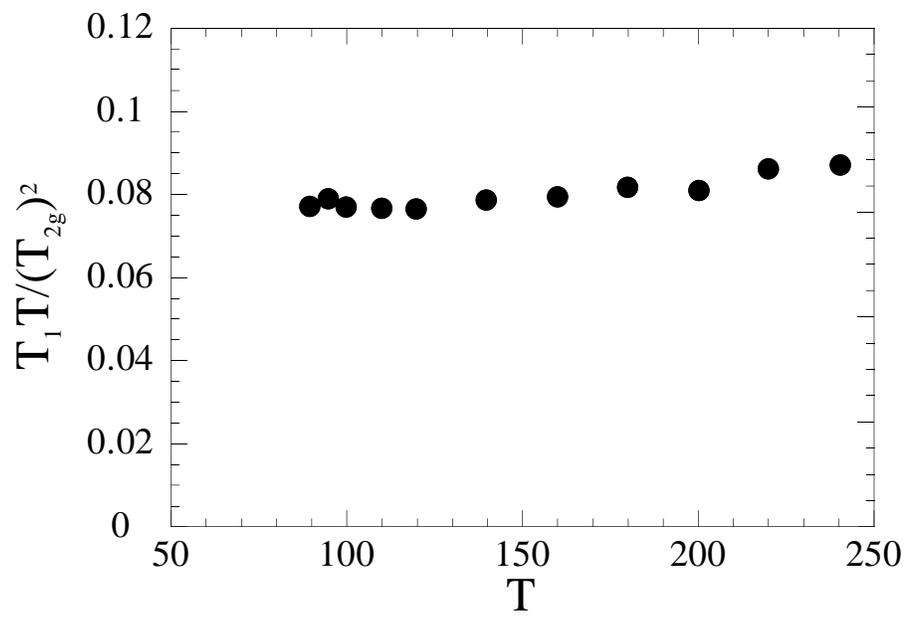
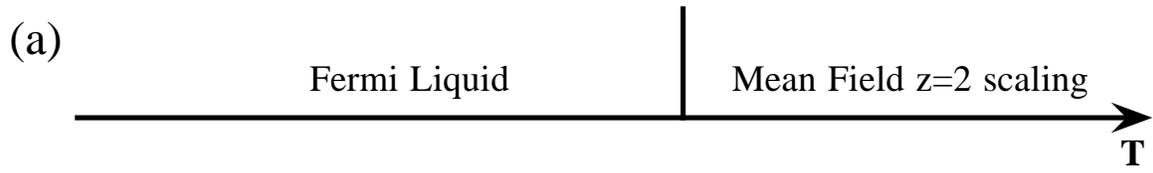
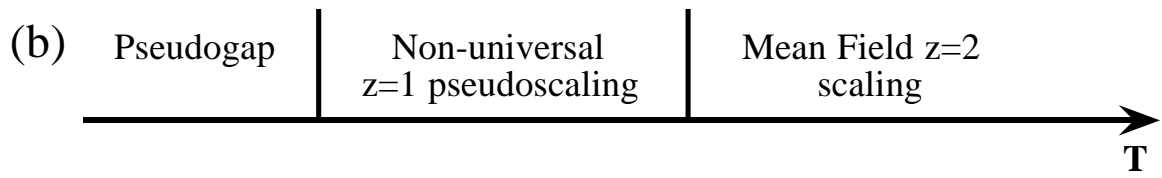
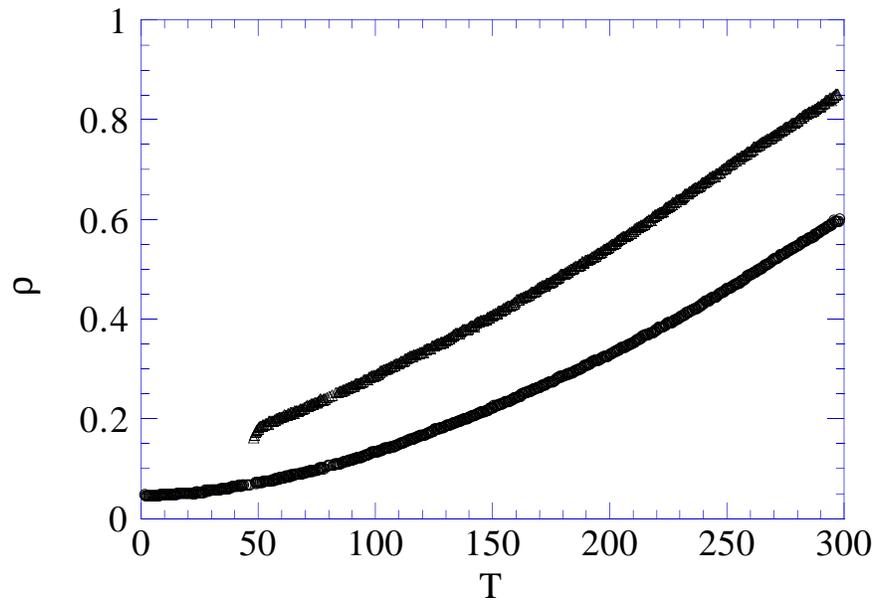
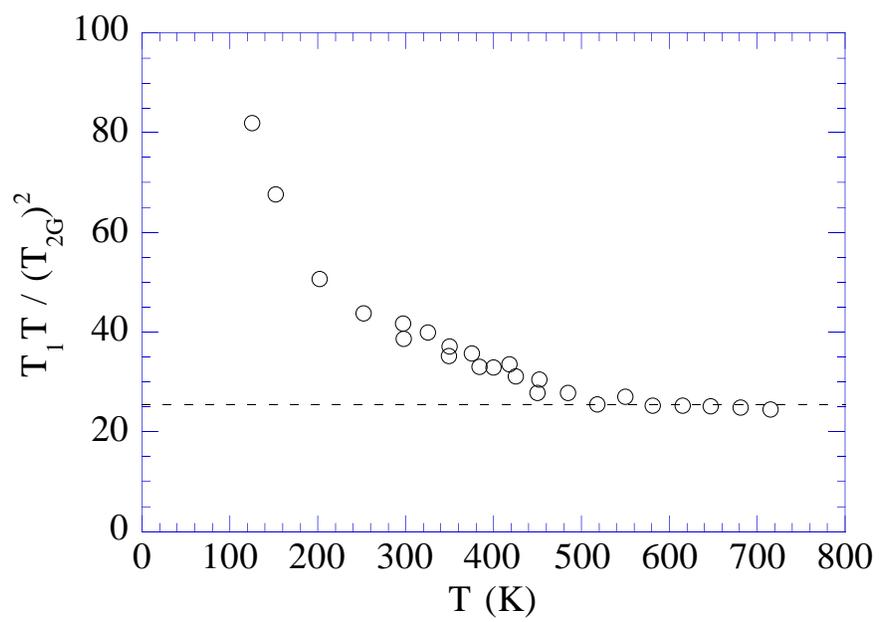
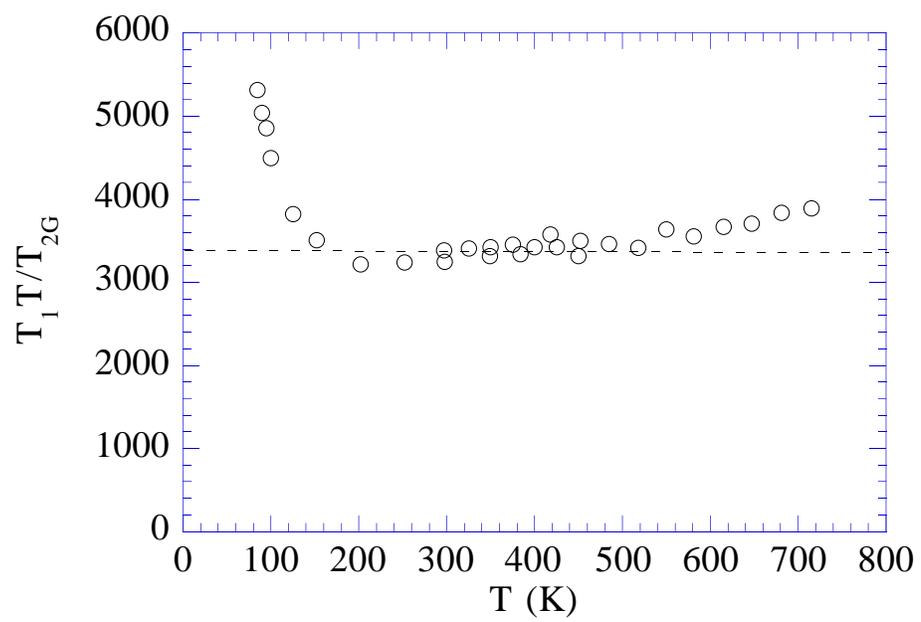
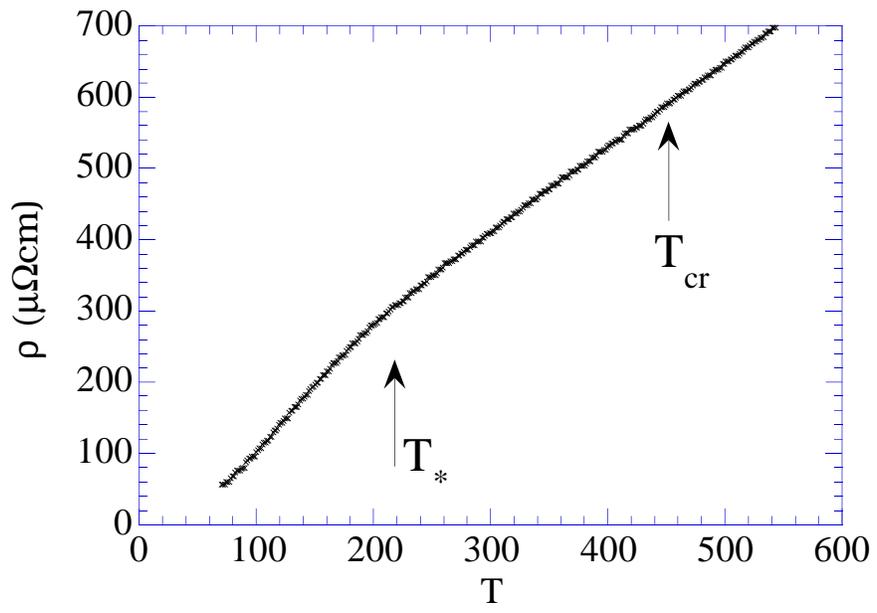
\end{document}